\def \cm{~\rm{cm}}
\def \s{~\rm{s}}
\def \km{~\rm{km}}

\def \K{~\rm{K}}

\def \AU{~\rm{AU}}
\def \erg{~\rm{erg}}

\def \yr{~\rm{yr}}

\documentclass[12pt,preprint]{aastex}

\shorttitle{VLTPs influenced by accretion} \shortauthors{Frankowski \& Soker}

\begin{document}

\title{VERY LATE THERMAL PULSES INFLUENCED BY ACCRETION IN PLANETARY NEBULAE}

\author{ Adam Frankowski\altaffilmark{1} and Noam Soker\altaffilmark{1}
}
\altaffiltext{1}{Department of Physics,
Technion$-$Israel Institute of Technology, Haifa 32000, Israel;
adamf@physics.technion.ac.il; soker@physics.technion.ac.il. }

\begin{abstract}
We consider the possibility that a mass of $\sim 10^{-5}-10^{-3}
M_\odot$ flows back from
the dense shell of planetary nebulae and is accreted by the central
star during the planetary nebula phase.
This backflowing mass is expected to have a significant specific angular
momentum even in
(rare) spherical planetary nebulae, such that a transient accretion disk
might be formed.
This mass might influence the occurrence and properties of a very late
thermal pulse (VLTP), and might even trigger it.
For example, the rapidly rotating outer layer, and the disk if still exist,
might lead to axisymmetrical mass ejection by the VLTP.
Unstable burning of accreted hydrogen might result in a mild flash of the
hydrogen shell, also accompanied by axisymmetrical ejection.
\end{abstract}

\section{INTRODUCTION}
\label{sec:intro}

A planetary nebula (PN) is an expanding ionized circumstellar cloud that
was ejected during the
asymptotic giant branch (AGB) phase of the stellar progenitor.
The nebula is ionized by the hot central star.
After several thousands years the nuclear reactions in the central star
cease, when more or less
that star starts its cooling track toward the the WD track (e.g., Kovetz
\& Harpaz  1981; Sch\"onberner 1981, 1983).
The PN shell is very dense, while the region between the central star
and the shell has a very low density.
One of the questions at the early history of PN modelling was why part
of the shell mass does not flow back to toward the central star?
This was answered with the suggestion (Kwok et al. 1978) and discovery
(Heap 1979; Cerruti-Sola \& Perinotto 1985) of a fast and tenuous wind blown by the
central stars of PNs.
The fast wind exerts pressure on the PN shell, mainly by forming a hot
low density bubble, accelerating the shell outward, and preventing
mass from flowing back.

However, under some circumstances small amounts of mass might flow back
toward the central star.
Accretion of backflowing material during the post-AGB phase was
mentioned before by Mathis \& Lamers (1992), Bujarrabal et al. (1998),
and Zijlstra et al. (2001), and studied by Soker (2001).
Most of these studies were interested in substantial mass, $\sim 0.001-0.1 M_\odot$,
backflowing from the nebular shell and accreted by the central star.
A much smaller mass, $\sim 10^{-6}  M_\odot$, is required
in the scenario, involving segregation of gas and dust in the outflow,
invoked by Mathis \& Lamers (1992) to explain chemical abundance
peculiarities observed in some post-AGB stars.

The backflowing mass might influence the evolution by forming accretion disk and
jets, and by supplying fresh material to the core and by that extending the
post-AGB (pre-PN) phase.
Soker (2001) showed that such a process requires that some of the
material ejected during the AGB phase has a very low outflow speed
$\la 1 \km \s^{-1}$, such that it can slow down and fall within a
few thousand years.
According to Soker (2001), the back flowing material falls because of
the gravitational attraction of the central star (or binary system),
and falls from a distance of $\sim 400 \AU$.
A binary interaction can lead to such an outflow.

In the present paper we consider a different kind of backflow, where
gravity is not important.
We consider a relatively small mass, $\sim 10^{-5}-10^{-3} M_\odot$, that
falls back during the PN phase rather than the pre-PN phase, and is
pushed back to the center by the thermal pressure of the ionized shell
rather than by gravity.
This is discussed in section \ref{sec:fallback}.

The small amount of mass that falls on the central star at the PN phase
might influence a very late thermal pulse (VLTP), or even trigger it.
Hajduk et al. (2007) considered the possibility that mass accretion can
induce a VLTP. They discuss this for the `old nova' CK Vul, where the mass is assumed
to come from a companion.
We instead, discuss the case where the accreted mass originates in the
nebula (Sect.~\ref{sec:fallback}).

A VLTP is a helium shell flash occurring so late in the post-AGB evolution that
the hydrogen burning shell is already extinct
(e.g., Fujimoto 1977; Iben et al. 1983; Herwig et al. 1999).
This allows the H-rich matter to be ingested directly into the convective
flash zone and leads to a rapid, looped evolution on H-R diagram (Herwig et al. 1999).
Other ``late'' thermal pulse scenarios, occurring earlier in the course of the
post-AGB evolution, before complete exhaustion of the H-burning shell, are
also considered in Sect.~\ref{sec:pulse},
but it appears that VLTP pertains the most to our fall-back scenario.
The possible influence of the accreted mass on the VLTP is discussed in
section \ref{sec:pulse}.
A short summary follows in section \ref{sec:summary}.

\section{BACKFLOW DURING THE PLANETARY NEBULA PHASE}
\label{sec:fallback}

\subsection{Formation of backflowing blobs}

We envision the following process that might lead to the formation of
blobs that
flow toward the central star.

The initial interaction of the fast wind blown by the post-AGB star with
the dense
shell blown during the end of the AGB phase is
momentum-driven, since the shocked fast wind material cools very fast
and the
acceleration of the shell is determined by momentum conservation along
each direction.
When the fast wind speed is high enough the post shock radiative cooling
time is longer than
the flow time and a hot bubble is formed.
Radiative cooling rate is small, and most of the energy is retained in
the flow.
The shell is in the energy-driven phase; see early works by Volk \& Kwok
(1985) and Kahn \& Breitschwerdt (1990).
The transition from momentum-driven to energy-driven shell occurs when
the fast wind speed
is $v_f \sim 160 \km \s^{-1}$ (Kahn \& Breitschwerdt 1990).
The interaction is prone to instabilities: the thin shell instability
during the momentum-driven
phase (Dwarkadas \& Balick 1998), and the  Rayleigh-Taylor instabilities
during the energy-driven
phase  (Kahn \& Breitschwerdt 1990; Dwarkadas \& Balick 1998).

The energy-driven phase is likely to start before the ionization phase
starts, or at
about the same time.
To show this we notice the following relation between the escape speed
from the
central star $v_{\rm esc}$, the central star effective temperature
$T_{\ast}$,
its mass $M_{\ast}$, and its luminosity $L_\ast$
\begin{equation}
v_{\rm esc} = 170
\left(\frac {T_\ast}{1.5 \times 10^4 \K} \right)
\left(\frac {M_\ast}{0.6 M_\odot} \right)
\left(\frac {L_\ast}{3000 L_\odot} \right)^{-1/4}
\km \s^{-1}.
\label{teff}
\end{equation}
Assuming that the wind speed is $v_f \simeq v_{\rm esc}$, we find that
the transition to
the energy-driven shell occurring at $v_f \simeq 160-200 \km \s^{-1}$
(Kahn \& Breitschwerdt 1990),
happens at about the same time that ionization starts to be significant.
This consideration is in accord with the central star model of Perinotto
et al. (2004), who
used a star with luminosity of $L_\ast= 6300 L_\odot$, and where the
wind speed of
$v_f \simeq 200 \km \s^{-1}$ is reached when the central
star temperature is $T_{\ast} = 1.6 \times 10^4 \K$.

The order of events implies that the hot bubble develops before the
dense shell of AGB wind is fully ionized, and with it the instabilities
(Kahn \& Breitschwerdt 1990; Dwarkadas \& Balick 1998).
Before ionization starts the pressure in the dense shell is of the order
of the pressure in the hot bubble.
When ionization becomes important it rapidly propagates through the
shell, and heats it to a temperature of $\sim 10^4 \K$ (Perinotto et al. 2004).
The shell pressure increases because both temperature and the number
density of particles increase.
The temperature increases from several$\times 100 \K$ to $\sim 10^4 \K$,
so we can take a pressure jump of $ \sim 100$.
The shell starts to expand into the hot bubble.
Because the initial pressure of the ionized shell is much larger than
that of the counter pressure
of the hot bubble, the front of the shell expands at a velocity
$v_{\rm exp}$ relative to the shell given by (Landau \& Lifshits 1987, $\S 99$)
$v_{\rm exp} \simeq {2 c_0}/({\gamma-1})$, where $\gamma$ is the
adiabatic exponent and $c_0 \simeq 10 \km \s^{-1}$ is the sound speed in
the ionized shell.
For $\gamma= 5/3$ and $\gamma= 1.1$ one finds $v_{\rm exp}= 3 c_0$
and $v_{\rm exp}= 20 c_0$, respectively.

The ionizing radiation will keep the inflowing mass at a temperature of
$\sim 10^4 \K$;
a value of $\gamma$ close to unity is appropriate for the flow.
The front of the flow does not contain much mass, and therefore the
relevant flow speed is a lower velocity, say
$(0.1-0.5)\times v_{\rm exp} = (0.2-1)\times c_0/({\gamma-1}) \simeq 20-100 \km \s^{-1}$,
where we took $\gamma\simeq 1.1$.
However, the acceleration will be less efficient than the one
dimensional flow derived in Landau \&Lifshits (1987, $\S 99$), because any blob formed by
instability will expand to the sides (transversely) as well.

In any case, we conclude that dense blobs formed by instability will
start to flow inward at a speed of $\sim 10-50 \km \s^{-1}$.
This speed is larger than the expansion velocity of the shell $\sim 10-20 \km \s^{-1}$,
and there will be a backflow at speed $v_{\rm back} \sim 3-30 \km \s^{-1}$ made up of blobs.
In the 1D calculation of Perinotto et al. (2004) the initial expansion
velocity of the shell is $10 \km \s^{-1}$.
After ionization starts a zone to the main shell is formed
with an expansion velocity of only $\sim 4 \km \s^{-1}$.
Namely, the inner zone of the main shell was substantially slowed down.
The pressure of the hot bubble prevents further acceleration inward,
and later accelerates the entire shell outward.
However, if a blob is disconnected from the main shell,
it will have the hot bubble material all around it.
The pressure of the hot bubble is more or less isotropic and it will not
slow the blob down. This effect cannot be studied with a 1D code.

The end product of the initial evolution is a large number of dense
(relative to the hot bubble) blobs at a temperature of $\sim 10^4 \K$ that are immersed
in the hot low density gas of the hot bubble.
Many of these will be flowing inward due to inward acceleration by the pressure of the
ionized shell during the early ionization stage.
If a blob does not slow down much, its backflow time from an initial
radius $r_0$ is
\begin{equation}
\tau_{\rm flow} \simeq \frac {r_0}{v_b} = 3000
 \left( \frac {r_0}{10^{17} \cm} \right)
\left(\frac {v_b}{10 \km \s^{-1}} \right)^{-1} \yr.
\label{tauflow}
\end{equation}
This is about the time required for the mass to be accreted during the PN
phase, but before a VLTP occurs.

\subsection{The properties of the blobs}
\label{Sect:blob_properties}

For the blob not to slow much it should encounter a mass $M_e$ not much
larger than
its initial mass $M_b= 4 \pi a_b^3 \rho_b/3$, where $a_b$ is the radius of
the blob and $\rho_b$ is its density.
For simplicity we assume the blob is in the shape of a sphere.
After flowing a distance $\sim r_0$ the blob encounters a mass
\begin{equation}
M_e \simeq  \pi a_b^2 r_0 \rho_h,
\label{me1}
\end{equation}
and the condition reads
\begin{equation}
1 \la   \frac {M_b}{M_e} \simeq  400 \left(\frac {a_b}{r_0}\right)
\left(\frac {\rho_b}{300 \rho_h} \right).
\label{me2}
\end{equation}
The density ratio of $300$ comes from pressure equilibrium between the
hot bubble
and the blob, and a temperature ratio of $\sim 3 \times 10^6 \K /10^4 \K
= 300$.
The blobs' radius is therefore $a_b \ga 0.0025 r_0$
The corresponding mass of the blob is
\begin{equation}
M_b \sim M_{\rm neb} \left( \frac{a_b}{r_0} \right)^{3} = 10^{-7}
\left(\frac {a_b}{0.01r_0} \right)^3
\left(\frac {M_{\rm neb}}{0.1 M_\odot} \right) M_\odot
\label{mb1}
\end{equation}
where $M_{\rm neb}$ is the mass in the nebula, and in the first equality
we
assume that the blob density is about equal to the density in the
nebula.

To survive, the blob must be thermally isolated from the surrounding
(otherwise it will be evaporated). We assume that tangled magnetic
fields
within the blob isolate the surviving blobs.
We cannot estimate the number of surviving blobs, but the requirements
of
($i$) an initial perturbation that will form a blob with negative radial
velocity
of $v_b \sim 10 \km \s^{-1}$, ($ii$) a large enough blob to maintain its
speed,
and ($iii$) suppression of heat conduction by magnetic fields, suggest
that their number will be limited.
We scale the number of blobs by $N_b = 1000$.

\subsection{Constraints on the fast wind evolution}
\label{Sect:fast_wind}

As long as the dense blob flows inward inside the hot bubble it does not
feel the
ram pressure of the fast wind.
When it gets out of the hot bubble through the inner boundary (reverse
shock),
the fast wind hits the blob directly, exerting a force of
$F_{\rm ram}= \rho_f v_f^2 \pi a_b^2$, where $\rho_f$ is the density of
the fast wind.
The deceleration is $a_{\rm ram}= F_{\rm ram}/M_b$. Using this value we
find the distance over which the
blob will stop, $r_{\rm stop} \simeq v_b^2/2 a_{\rm ram}$,  to be
\begin{equation}
r_{\rm stop} \simeq \frac{2}{3} \left( \frac{v_b}{v_f} \right)^2
\left(\frac {\rho_b}{\rho_f} \right) a_b .
\label{rstop1}
\end{equation}
While inside the hot bubble the density in the blob is determined by
pressure equilibrium with the hot
bubble.
Close to the inner boundary of the hot bubble, $r_{\rm bubble}$,  the
pressure is about
equal to the ram pressure of the fast wind there, and so is the blob
pressure.
This gives $\rho_b \simeq \rho_f (v_f/c_0)^2$, where $c_0 \simeq 10 \km
\s^{-1}$
is the sound speed in the blob.
In our scenario $c_0 \simeq {v_b}$, and we find from equation
(\ref{rstop1}) the condition
for the blob to reach the center to be $a_b \ga r_{\rm bubble}$.
Namely, the blob must be large when it gets out of the hot bubble; we'll
come
back to this point later.

Equating the ram pressure to the pressure in the blob $\rho_f v_f^2
\simeq \rho_b c_0^2$, and
substituting for the fast wind density $\rho_f = \dot M_f/(4 \pi r^2
v_f)$, and for the
density in the blob $M_b/(4 \pi a_b^3/3)$, we can isolate $a_b$.
Inserting this expression
for the blob radius $a_b$ in the condition $a_b \ga r_{\rm bubble}$ we
derive the condition
\begin{equation}
r_{\rm bubble} \la 3 \frac {M_b c_0^2}{\dot M_f v_f} = 2.5 \times
10^{15} 
\left( \frac {M_b}{10^{-7} M_\odot} \right)
\left( \frac {\dot M_f}{10^{-11} M_\odot \yr^{-1}} \right)^{-1}
\left( \frac {v_f}{4000 \km \s^{-1}} \right)^{-1} \cm,
\label{rstop2}
\end{equation}
where in the second equality we have substituted $c_0= 10 \km \s^{-1}$.

When the fast wind properties are constant with time, the inner shock
(the inner boundary
of the hot bubble $r_{\rm bubble}$) moves outward as the nebula expands.
We require, therefore, that within a short time the fast wind momentum
discharge,
defined as ${\dot M_f v_f}$, will decrease by a large factor. In that
case
$r_{\rm bubble} \propto ({\dot M_f v_f})^{1/2} $ .
In the model of Perinotto et al. (2004), $\dot M_f$ drops from $\sim
10^{-8} M_\odot$
to $\sim 10^{-10} M_\odot \yr^{-1}$ within $\sim 1000 \yr$ at an age of
$t\simeq 7000 \yr$.
The fast wind speed then is $\sim 10^4 \km \s^{-1}$.
Taking values for the inner hot bubble boundary from their model, as
well as other parameters,
we crudely scale the parameters as
\begin{equation}
r_{\rm bubble} \sim 2 \times 10^{15}
\left( \frac {\dot M_f}{10^{-11} M_\odot \yr^{-1}} \right)^{1/2}
\left( \frac {v_f}{4000 \km \s^{-1}} \right)^{1/2} \cm.
\label{rstop3}
\end{equation}
Substituting this crude relation in equation (\ref{rstop2}) we find for
our
demand for mass accretion to occur
\begin{equation}
\left( \frac {\dot M_f}{10^{-11} M_\odot \yr^{-1}} \right)
\left( \frac {v_f}{4000 \km \s^{-1}} \right) \la
\left( \frac {M_b}{10^{-7} M_\odot} \right)^{2/3}.
\label{rstop4}
\end{equation}

The condition $a_b \ga r_{\rm bubble}$  for the mass to reach the center
as derived
above, implies that the backflowing gas covers a large solid angle.
Let the solid angle covered by the backflowing gas be $4 \pi \beta$.
The ratio of the force due to the ram pressure, $F_f= \beta \dot M_f
v_f$ of the fast
wind to the gravitational force of the central star $F_G= G M_\ast M_b
/r^2$ is
\begin{equation}
\frac{F_f}{F_g} =
\left( \frac {r}{5 \times 10^{14} \cm} \right)^2
\left( \frac {\dot M_f}{10^{-11} M_\odot \yr^{-1}} \right)
\left( \frac {v_f}{4000 \km \s^{-1}} \right)
\left( \frac {M_b}{10^{-7} M_\odot} \right)^{-1}
\left( \frac {M_\ast}{0.6 M_\odot} \right)^{-1}
\left( \frac {\beta}{0.25} \right).
\label{force1}
\end{equation}

~From the derivations above we learn the following.
Equations (\ref{rstop4}) shows that for the backflowing blob to reach
the center the
fast wind should be extremely weak, e.g., the mass loss rate is an order
of magnitude below
that in the model of Perinotto et al. (2004) at the age of 8000 years.
Considering the many unknown in the evolution of the central stars of
PNs, we consider this
constraint on the fast wind not unreasonable. Namely, we conjecture that it
applies in many cases at the age of several thousand years. From equation (\ref{force1})
it turns out that gravity becomes important
when the backflowing warm gas reaches a distance of $\sim 100 \AU$
under these conditions.
The constraint on the blob mass is $M_b \ga 10^{-7}- 10^{-6} M_\odot$.

\subsection{Angular momentum}
\label{Sect:angular_momentum}

\subsubsection{Single central star}

In section 2.1 we considered the radial acceleration by the nebular
thermal pressure
of small amount of mass toward the center.
Let us assume that because of stochastic motion in the unstable layer
the pressure
gradient exerts a small azimuthal (tangential) acceleration in addition
to the
radial acceleration inward.
Let the typical azimuthal velocity be $\delta v_b \sim 0.01
(\delta/0.001) \km \s^{-1}$,
such that the typical angular momentum is $j_b= \delta v_b r_0$.
The combined specific angular momentum of $N_b$ blobs with randomly
oriented
angular momentum is $j \simeq N_b^{-1/2} j_b$.
This angular momentum corresponds to a Keplerian disk at a radius $r_d$
around the
central star of mass $M_\ast$ given by
\begin{equation}
r_d \simeq 10^{11}
\left(\frac {N_b}{1000} \right)^{-1}
\left(\frac {\delta}{0.001} \right)^2
\left(\frac {v_b}{10 \km \s^{-1}} \right)^2
\left(\frac {r_0}{10^{17} \cm} \right)^2
\left(\frac {M_\ast}{0.6 M_\odot} \right)^{-1} \cm.
\label{rd1}
\end{equation}
This radius is larger than the radius of the central star during the PN
phase, e.g.,
the VLTP occurs when the radius of the central star might be as small as
$\sim 10^9 \cm$
(e.g., Iben et al. 1983).
We conclude that an accretion disk might be formed around the central
star.
The accretion is likely to occur over hundreds to thousands of years, with an average
mass accretion rate
of $\dot M_{\rm acc} \sim 10^{-9}-10^{-6} M_\odot \yr^{-1}$.
Jets might be launched.
In any case, if the disk exists during the VLTP, the
mass ejection, perhaps resembling a nova eruption,
will be axisymmetrical.
If the PN is not spherical, then the nebula itself might
have some amount of initial angular momentum, making an accretion disk
formation more likely.

\subsubsection {Interaction with a wide companion}

An interaction with a wide companion is another process by which the
backflowing material can acquire angular momentum.
We will take the companion to be of low mass, e.g., a low mass main sequence
star, a brown dwarf, or a massive planet.
This process is more general, and can be relevant to inflow from smaller
radii, and at earlier phases of evolution.
For example, during the final AGB and early post-AGB phases if mass falls
back from the 'effervescent zone' (Soker 2008).
The effervescent zone is an extended zone conjectured to exist above evolved
AGB stars (those with high mass loss rates) extending to $\sim 100-300 \AU$.
In addition to the escaping wind, in this zone there are parcels of
gas that do not reach the escape velocity. These parcels of dense gas rise
slowly and then fall back.

Let the companion orbit the central star at radius $a$.
Gas falling to the center from radius $r_0>a$
will have an infall velocity of
\begin{equation}
v_{\rm in} = \left[ \frac {2G M_\ast}{a} \left( 1-\frac {a}{r_0} \right)
\right]^{1/2}.
\label{vin1}
\end{equation}
The orbital velocity of the companion is $v_2=(G M_\ast/{a})^{1/2}$.
The relative velocity of the inflowing gas and the companion
is given by $v_{\rm rel}^2=v_2^2+v_{\rm in}^2 \simeq {2G M_\ast}/{a}$,
where we assume that $r_0 >1.5 a$.
The accretion radius of the companion is the Bondi-Hoyle-Lyttleton one
\begin{equation}
R_{a2} = \frac {2G M_2}{v_{\rm rel}^2} \simeq
\frac{M_2}{M_\ast}a,
\label{racc1}
\end{equation}
where $M_2$ is the mass of the companion.

The mass with an impact parameter of $b \la R_{a2}$ is accreted by the
companion.
Mass at larger distances applies drag force on the secondary, with a
magnitude of (Alexander et al. 1976)
\begin{equation}
F=\pi R_{a2}^2 \rho v_{\rm rel}^2 \ln (R_{\rm max}/R_{a2}).
\label{f1}
\end{equation}
The same force is applied by the star on the gas.
The azimuthal component of this force is $F_{\theta} = Fv_2/v_{\rm rel}$.
We also take the maximum radius of influence to be $R_{\rm max}\simeq a$,
which by equation (\ref{racc1}) gives
$\ln (R_{\rm max}/R_{a2}) \simeq \ln(M_\ast/{M_2}) \simeq 1$, for $M_2 \simeq 0.1 M_\ast \simeq 0.1 M_\odot$.

The angular momentum imparted to the inflowing gas by this gravitational
interaction is
\begin{equation}
\frac{dJ}{dt} = F_\theta a \simeq   \pi R_{a2}^2 \rho v_{\rm rel} v_2 a .
\label{jt1}
\end{equation}
The total inflow rate through radius $a$ is
$\dot M_{\rm in} = 4 \pi a^2 \rho v_{in}$.
The specific angular momentum of the inflowing gas is
$j=({dJ}/{dt})/\dot M_{\rm in}$.
Using our approximations, the value of $j$ is
\begin{equation}
j_{\rm in}  \simeq \left( \frac{M_2}{M_\ast} \right)^2 v_2 a .
\label{j1}
\end{equation}
To examine the possibility of disk formation around the central star, $j_{\rm in}$
should be compared with the specific angular momentum of a Keplerian circular
orbit on the star equator $j_\ast=(G M_\ast R_\ast)^{1/2}$.
We find, after using the expression for $v_2$,
\begin{equation}
\frac{j_{\rm in}}{j_\ast} \simeq \left( \frac{M_2}{M_\ast} \right)^2
\left( \frac{a}{R_\ast} \right)^{1/2} \simeq
\left( \frac{M_2}{0.1 M_\ast} \right)^2
\left( \frac{a}{100 \AU} \right)^{1/2}
\left( \frac{R_\ast}{1 R_\odot} \right)^{-1/2}
\label{jj}
\end{equation}
where in the second equality we have substituted values appropriate for a
central star of a PN in our studied case of a wide low mass companion.

We see that a binary companion can spin-up the inflowing gas to a degree
that will lead to the formation of a disk.
If this occurs at early stages of the post-AGB phase, it can lead to the
formation of accretion disk that launches two jets, and by that shapes
the PNs.
We note that the low mass companion, $M_2 \simeq 0.1 M_\ast$, will not
accrete much mass by itself.

To summarize this entire section, we did not demonstrate that a backflow
forms in the PN phase;
this will require intensive 3D numerical simulations, because the
dynamical
range from perturbations to a backflow to the center is very large.
We only showed here that it is quite plausible that dense blobs
will flow from the nebula back to the central star, and will cause the
star outer layers to rotate fast; it might even form an accretion disk.
These effects will result in an axisymmetrical mass ejection if a VLTP
take place.

\section{THE INFLUENCE OF ACCRETION ON A VERY LATE THERMAL PULSE}
\label{sec:pulse}

Various objects have their appearance (in particular, chemical
abundances) explained as due to a final thermal pulse occuring
during the transition from the AGB to the PN phase.
To name a few examples (see e.g.~recent reviews by
Werner \& Herwing 2006 and Kimeswenger et al. 2008):
FG Sagittae (the prototypical born-again AGB object),
Abell 30, Abell 78, CK Vul,
PG 1159 (and the class of pulsating white dwarfs named after it),
and WR-like central stars of PNe.
If the above description of PN fall-back is correct,
it is worth checking if a connection can be drawn between
the fall-back and VLTP.

In general, three types of ``late'' thermal pulse scenarios can be distinguished
(e.g., Herwig et al. 1999, Bl\"ocker 2003).
(1) AGB Final Thermal Pulse (AFTP) takes place at the very end of the AGB
phase, right before the star leaves this evolutionary stage.
(2) Late Thermal Pulse (LTP) occurs when a star evolves off the AGB at phase
$\gtrsim 0.85$ of the thermal pulse cycle and it is caught by a TP
on the horizontal part of its post-AGB evolution in the H-R diagram.
(3) Very Late Thermal Pulse occurs still later, when the hydrogen burning shell
is extinguished and a star is descending along the white dwarf cooling
track towards lower luminosities.
In the VLTP case, the H-rich envelope is penetrated by the pulse-driven
convective zone which leads to an additional short-lived H-flash.
The abundance patterns produced in these three cases are somewhat different,
but they all produce H-deficient ejecta and H-deficient post-AGB objects.

Can the accretion of back-falling PN
matter, as described in the previous section, affect when and how a
``late'' TP occurs?
By definition, it cannot affect an AFTP (no PN yet at this stage),
and interfering with an LTP seems precluded by the requirement that
the fast wind should practically cease for fall-back to become possible
(Sect.~\ref{Sect:fast_wind}). That does not happen before moving below the
high temperature knee on the HR diagram (e.g., Perinotto et al. 2004).
Unlike for gravity driven backflows considered by Soker (2001, 2008),
this late PN backflow cannot prolong the post-AGB lifetime at the horizontal
part of the track and increase the chances for an LTP (as proposed in
Zijlstra et al. 2001 for CK Vul).
Note, however, that many main sequence late O dwarfs seem to have winds
about two orders of magnitude weaker than expected from radiative winds
theory (see Marcolino et al. 2009 for a recent update on the topic).
If a similar phenomenon would exist among post-AGB stars, it would make fall back
to the central star more likely, be it before or after the hight temperature knee in HR diagram.
We will now focus on the VLTP.

According to sections~\ref{Sect:blob_properties} and \ref{Sect:fast_wind},
fall-back would add $10^{-5}-10^{-3}M_{\odot}$ of H-rich matter to the
envelope of a PN central star before a VLTP occurs.
Burning of this material can be either stable on unstable, depending on the
exact conditions (accretion rate, metallicity, central star mass and
intrinsic luminosity, see Shen \& Bildsten 2007, esp. their Fig.~3).
The relatively high intrinsic luminosity of the young pre-WD (a few hundred
$L_\odot$) has a stabilising effect on H-shell burning and should
be taken into account in the stability considerations,
unlike in classical novae.
In the $0.6 M_{\odot}$ core mass model used by Perinotto et al. (2004),
at the time when fall-back is expected to occur, i.e, at the age of
$8000-10\,000$yrs, the intrinsic luminosity is $\sim 10^{36} \erg \s^{-1}$.
According to the results of Shen \& Bildsten (2007), this leads to
unstable H-shell burning only in a narrow range of accretion rates,
$0.3-6 \times 10^{-8} M_{\odot} \yr^{-1}$.

In the case of a stable H-shell burning,
the effect of the accreted matter would mainly be to prolong the residual
H-shell burning and keep the star longer in the evolutionary stage not far
below the high-temperature knee on the H-R diagram.
Accreting $10^{-5}-10^{-3} M_{\odot}$ of H-rich matter on a
$0.6 M_{\odot}$ pre-WD at a stable-burning rate of $\sim 10^{-9} M_{\odot}$
(Shen \& Bildsten 2007) would take $10^{4}-10^{6}$ yrs. This is
comparable to the duration of a TP cycle ($\sim 10^{5}$ yrs for a
$0.6 M_{\odot}$ core mass) and processes a similar amount of matter
(an AGB star processes $10^{-2}-10^{-3} M_{\odot}$ between two consecutive
TPs, with the higher values characterizing lower mass cores, $\lesssim 0.6
M_{\odot}$).
Effective fall-back would therefore increase the number
of post-AGB objects experiencing a VLTP above the standard
estimate of $\sim 10\%$ (Iben 1984). The increase would be
more significant for the more massive pre-WD objects,
as they require less mass processed and accumulated in the
He-shell for a TP to occur.
Notice also that after accretion the star will rotate due to the angular
momentum transferred, which may affect the flash intrinsic geometry
(in addition to constrainig the ejecta, as described in
Sect.~\ref{Sect:angular_momentum})
and possibly other characteristics. This is different
from ``normal'' late TPs, in which the core is rather not expected
to rotate rapidly.

On the other hand, if the fall-back accretion process is
slow enough (but not as slow as to be stabilized by the residual core luminosity),
H-shell burning becomes unstable.
In this case, fall-back would also switch on a thermonuclear
instability on the pre-WD, but in the form of H-shell flashes. They would be
similar to weak nova eruptions, with the H-shell igniting rather than the
He-shell (as is in a VLTP). However, from the external appearance -- new ejecta
within an old PN -- they could easily be taken for a VLTP.
In the past, Iben \& MacDonald (1986) introduced a term {\em self-induced nova}
for a final flash of H-burning they found to happen late in the evolution of
a cooling WD with a very thin He layer buffering between the CO core and the
H-rich envelope. In their case the ignition was due to diffusive mixing
of C and H through this thin buffer and occured very late in the WD cooling
stage ($2\times 10^{7}$yrs after the PN stage, when the luminosity of their
model dropped to $0.3 L_\odot$).
The fall-back induced flash described here would also be a
{\em self-induced nova}, but of a distinct kind, less violent and much earlier
in the evolution, when the pre-WD is still hot and luminous and its ejected PN
is still around.

\section{SUMMARY}
\label{sec:summary}

We have presented a mechanism for a backflow of matter to occur
relatively late during PN evolution.
The proposed fall-back consists of dense blobs that are formed from instabilities at
the discontinuity between the slow and the fast winds of the central star.
They detach from the dense shell of the slow AGB wind at the beginning of the PN
ionization stage and fall back into the hot bubble with a velocity of several $\km \s^{-1}$.
They can reach the PN central star only when the fast wind mass loss rate drops to
$\la 10^{-11} M_\odot \yr^{-1}$.

The accreted mass is likely to posses enough angular momentum to spin-up the central star, and
even to form an accretion disk.
The estimated accretion rates of
$\dot M_{\rm acc} \sim 10^{-9}-10^{-6} M_\odot \yr^{-1}$, lasting for
hundreds to thousands of years, may lead to the formation of an accretion
disk and to launching of jets, even in the case of single central stars
of PNe.
The accreted mass, $\sim 10^{-5}-10^{-3} M_\odot$, would be enough to
increase the chance of a VLTP or even induce an H-shell flash,
leading to a mild {\em self-induced nova} within a PN.
The VLTP (or H-flash) ejecta would exhibit bipolar morphology, offering a
binary-independent explanation of point-symmetric inner nebular structures
prevalent among purported VLTP objects.

The estimates provided in this paper do not constitute a proof of
the existence of such PN fall-back, but they show that it is a reasonable
possibility. Numerical simulations of 3D radiative hydrodynamics, detailed
enough to resolve the blobs, will be required for a definite statement.
In any case, this scenario, together with other types of post-AGB backflow
(as described by Soker 2001 and 2008), serves to extend the theoretical
argument for the reality and importance of backflow in the PN phase.

This research was supported by the Asher Space Research Institute in the
Technion and by the Israel Science foundation.


\begin{references}

\reference{} Alexander, M. E., Chau, W. Y., \&  Henriksen, R. N. 1976, ApJ,
204, 879

\reference{} Bl\"ocker, T. 2003, in: Planetary Nebulae: Their Evolution and
Role in the Universe, S. Kwok, M. Dopita, \& R. Sutherland (eds.), ASP, IAU
Symp. 209, 101

\reference{} Bujarrabal, V., Alcolea, J., \& Neri, R. 1998, ApJ, 504,
915

\reference{} Cerruti-Sola, M., \& Perinotto, M. 1985, ApJ, 291, 237

\reference{} Dwarkadas, V. V., \& Balick, B. 1998, ApJ, 497, 267

\reference{} Fujimoto, M. Y., 1977, PASJ, 29, 331

\reference{} Hajduk, M., Zijlstra, Albert A., van Hoof, P. A. M. et al.
2007, MNRAS, 378, 1298

\reference{} Heap, S. R. 1979, in: Mass loss and evolution of O-type stars,
P. S. Conti \& C. W. H. de Loore (eds.), D. Reidel, IAU Symp. 83, 99

\reference{} Herwig, F., Bl\"ocker, T., Langer, N., \& Driebe, T. 1999,
A\&A 349, L5

\reference{} Iben, I. Jr. 1984, ApJ, 277, 333

\reference{} Iben, I. Jr., \& MacDonald, J. 1986, ApJ, 301, 164

\reference{} Iben, I. Jr., Kaler, J. B., Truran, J. W., Renzini, A. 1983,
ApJ, 264, 605

\reference{} Kahn, F. D., \& Breitschwerdt, D. 1990, MNRAS, 242, 505

\reference{} Kimeswenger, S., Zijlstra, A. A., van Hoof, P. A. M.,
Hajduk, M., Herwig, F., Lechner, M. F. M., Eyres, S. P. S.,
van de Steene, G. C. 2008, in: Hydrogen-Deficient Stars,
K. Werner \& T. Rauch (eds.), ASP Conf. Ser. 391, 177

\reference{} Kovetz, A. \& Harpaz, A.  1981, A\&A, 95, 66

\reference{} Kwok, S., Purton, C. R., \& Fitzgerald, P. M. 1978, ApJ,
219, L125

\reference{} Landau, L. D. \& Lifshits, E. M. 1987, Fluid Mechanics (2nd
edition), (Butterworwh-Heinemann, Burlington, MA).

\reference{} Marcolino, W. L. F., Bouret, J.-C., Martins, F., Hillier, D. J.,
Lanz, T., \& Escolano, C. 2009 (arXiv:0902.1833M)

\reference{} Mathis, J. S. \& Lamers, H. J. G. L. M. 1992, A\&A, 259, L39

\reference{} Perinotto, M., Sch\"onberner, D., Steffen, M., \& C.
Calonaci, C. 2004, A\&A, 414, 993

\reference{} Sch\"onberner, D. 1981, A\&A, 103, 119

\reference{} Sch\"onberner, D. 1983, ApJ, 272, 708

\reference{} Shen, K. J., \& Bildsten, L. 2007, ApJ, 660, 1444

\reference{} Soker, N. 2001, MNRAS, 328, 1081

\reference{} Soker, N. 2008, NewA, 13, 491

\reference{} Volk, K., \& Kwok, S. 1985, A\&A, 153, 79

\reference{} Werner, K., \& Herwig, F. 2006, PASP, 118, 183

\reference{}  Zijlstra, A., A., Chapman, J. M., te Lintel Hekkert, P.,
Likkel, L.,
 Comeron, F., Norris, R. P., Molster, F. J., \& Cohen, R. J. 2001,
MNRAS, 322, 280


\end{references}
\end{document}